\documentclass[aps,prl,preprint,showpacs]{revtex4}

\usepackage{graphicx}
\usepackage{amsmath}

\begin{document}

\title{Similarity between positronium-atom and electron-atom scattering}

\author{I. I. Fabrikant}
\affiliation{Department of Physics and Astronomy, University of Nebraska,
Lincoln, Nebraska 68588-0299, USA}
\author{G. F. Gribakin}
\affiliation{School of Mathematics and Physics, Queen's University Belfast,
Belfast BT7 1NN, Northern Ireland, UK}

\date{\today}

\begin{abstract}
We employ the impulse approximation for description of positronium-atom 
scattering. Our analysis and calculations of Ps-Kr and Ps-Ar collisions
provide theoretical explanation of
the similarity between the cross sections for positronium scattering
and electron scattering for a range of atomic and molecular targets
observed by S.~J.~Brawley \textit{et al.} [Science {\bf 330}, 789 (2010)].
\end{abstract}


\maketitle

Most of the methods employed in the theory of leptonic and atomic collisions
are based on solving the Schr\"{o}dinger equation with inclusion of pair
lepton-lepton and lepton-nuclear interactions.
Although the interactions are well known, for complex projectiles and targets 
such an approach is computationally very involved. An alternative approach 
is based on many-body equations involving scattering amplitudes, for example,
Faddeev equations \cite{Fad61}. The advantage of this approach is in 
the possibility of 
using amplitudes representing highly correlated motion between a part of the
projectile and the target. A typical example is the collision of a 
highly-excited (Rydberg) atom with a ground-state atom or a molecule. In this 
problem the scattering amplitude can be expressed in terms of the amplitudes
for electron and ion-core scattering by the ground-state atom
\cite{deP83,Mat84}. This is the idea behind the impulse approximation
\cite{Che52}.

Another example is the positronium (Ps) scattering by neutral targets. The
Ps atom is easily ionized (i.e., broken up) above the ionization threshold
(6.8 eV), and in
fact the ionization of Ps is becoming the dominant process in Ps-atom collisions
at collision energies above about 20 eV \cite{Lar12}. The Ps ionization energy
is much smaller than the ionization energies of the noble-gas atoms and many
small molecules (such as H$_2$, N$_2$, O$_2$, CO$_2$ or SF$_6$). This allows us
to consider the Ps atom as a losely bound system and the Ps-atom (molecule)
scattering as a coherent superposition of $e^-$-atom and $e^+$-atom scattering
processes.

Recently observed similarities between the Ps scattering and the electron
scattering from a number of atoms and molecules \cite{Bra10,Bra10a} suggest
that both processes are largely controlled by the same interactions. When
plotted as a function of the projectile velocity, the electron and Ps cross
sections are very close and
even show similar resonance-like features. This seems strange at first since in
the electron-atom scattering the electrostatic and polarization forces play a
role, while both seem to be absent in the Ps-atom scattering. However, at
intermediate energies electron scattering by noble-gas atoms is dominated by
a strong exchange interaction. There is a range of energies above the
Ramsauer minimum for the $e^-$-atom scattering, where the polarization is
less significant, but the energy is still not too high, so that $e^-$ or $e^+$
Rutherford scattering is not dominant.
In this energy range electron scattering by atoms and molecules
is strongly affected by the
exchange interaction whereas the positron scattering is relatively weak because of the
mutual cancellation of the repulsive static and attractive polarization 
forces. As a result, in the intermediate energy range, typically 
between about 5 and 50 eV,
positron scattering cross sections are significantly smaller than their electron counterparts \cite{Kimura00,Sur05}.

Close-coupling calculations of Blackwood {\it et al.} \cite{Bla02} produced
total cross sections for Ps scattering by noble-gas atoms that are
substantially lower than the corresponding electron scattering cross sections,
and lie below the experimental values \cite{Lar12,Bra10,Bra10a}.
These calculations allowed for the distortion
and break-up of Ps, but they were performed in the frozen-target approximation,
i.e., they did not take into account virtual excitation of the target. In
low-energy electron- and positron-atom collisions such virtual excitations
can be described in terms of the polarization interaction, and are known to be
important. For low-energy Ps-atom collisions they give rise to the van der
Waals interaction. In the intermediate-energy range above the Ps excitation
threshold, which we are interested in, the static van der Waals interaction is
not appropriate for the description of dynamical correlations between the
Ps and the target. The most direct way to include them for Ps-atom scattering 
is by extending the close-coupling calculations to account for the virtual
excitations of the target. While such calculations have been performed for Ps
collisions with the H atom \cite{Cam98,Bla02a} and would be an ultimate
goal in the problem of Ps-atom collisions, in the present paper we demonstrate
that a much simpler method based on the impulse approximation can account for
dynamical correlations, at least in the intermediate energy range important
for the experiments \cite{Bra10,Bra10a,Gar00}. This method also offers a
theoretical explanations for the similarity between electron-atom and Ps-atom
scattering.

Consider the scattering process
\[
{\rm Ps}(a,{\bf p}_i)+ A\to {\rm Ps}(b,{\bf p}_f)+A,
\]
where $a$ and ${\bf p}_i$ denote the internal state and center-of-mass momentum
of the incident Ps, while $b$ and ${\bf p}_f$ are their values
in the final state.

\begin{figure}[ht!]
\centering
\includegraphics[width=12cm]{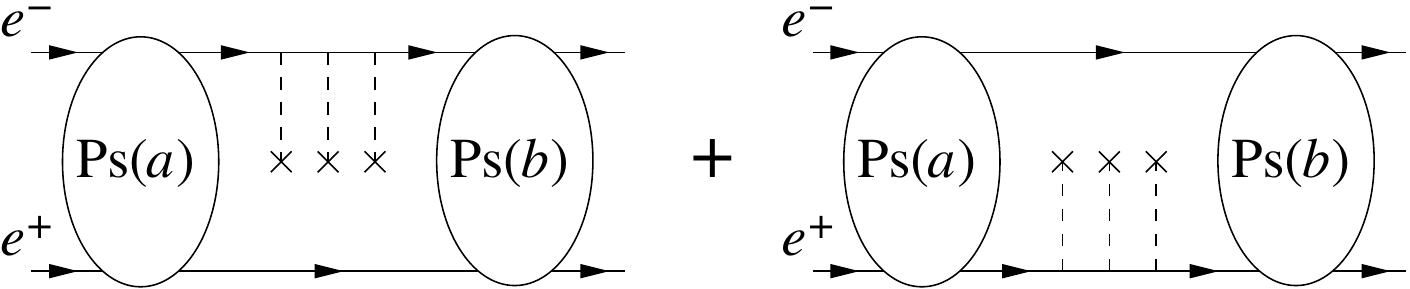}
\caption{Schematic diagrams of the approximation for the Ps-atom scattering
amplitude. The dashed lines with a cross show the interaction between
the electron or positron and the atom, which is included in all orders.}
\label{fig:Ps_diag}
\end{figure}

Compared with noble-gas atoms, the Ps is a diffuse and weakly-bound system.
Consequently, we can assume that when the Ps is scattered off such
targets, the Coulomb interaction within the Ps atom is weak in comparison
with the electron-atom or positron-atom interactions. In this case
the scattering amplitude can be approximated by the sum of two contributions
shown schematically in Fig.~\ref{fig:Ps_diag}.

Each of the two diagrams in Fig.~\ref{fig:Ps_diag} is in fact a
perturbation-theory sum which includes the electron-atom or positron-atom
interaction to all orders. Owing to the diffuse nature of Ps and low relative
velocities inside Ps, the particle which does not interact
with the target (i.e., the positron in the first diagram, and the electron in
the second diagram in Fig.~\ref{fig:Ps_diag}) does not change its
instantaneous momentum. 
We assume that the state of atom $A$ does not change during the collision.
However, the virtual excitations of the target are accounted for implicitly
in the electron and positron scattering amplitudes if they are calculated
beyond the static (or static-exchange) approximation. As a result, the Ps-atom
scattering amplitude can be written as the sum of two terms \cite{Mat84,Sta05}
(in atomic units),
\begin{align}\label{eq:ampl}
f_{ba}({\bf p}_f,{\bf p}_i)&= 2\int g^*_{b}({\bf q})
f^-({\bf v}^-_f,{\bf v}^-_i)
g_a({\bf q}+\Delta{\bf p}/2){d^3{\bf q}} \nonumber \\
&+2\int g^*_{b}({\bf q})
f^+({\bf v}^+_f,{\bf v}^+_i)
g_a({\bf q}-\Delta{\bf p}/2){d^3{\bf q}} 
\end{align}
where $\Delta{\bf p} ={\bf p}_f-{\bf p}_i$ is the change in the Ps momentum,
$g_a({\bf q})=(2\pi)^{-3/2}\int e^{i{\bf q}\cdot {\bf r}}\varphi _a({\bf r})
d^3{\bf r}$ is the Ps internal wave function in the momentum space
(${\bf r}={\bf r}_{e^+}-{\bf r}_{e^-}$ being the relative position vector), the
factor 2 is due to the Ps mass, and $f^\pm ({\bf v}',{\bf v})$
are the positron-atom and electron-atom scattering amplitudes
for the initial and final velocities 
\begin{equation}\label{eq:vv}
{\bf v}_i^\pm ={\bf p}_i/2 -\Delta{\bf p}/2\pm{\bf q},~~
{\bf v}_f^\pm ={\bf p}_i/2 +\Delta{\bf p}/2\pm{\bf q}.
\end{equation}
The amplitudes $f^\pm$ in Eq.~(\ref{eq:ampl}) are off the energy shell in the
sense that $|{\bf v}_i^\pm|\neq |{\bf v}_f^\pm |$. Besides, each amplitude
depends on the energy argument
$E^\pm ={\bf p}_i^2/4+\varepsilon _a-|{\bf v}_i^\pm |^2/2$ not equal to physical 
energy $|{\bf v}_i^\pm |^2/2$ \cite{deP83,Fab92}. In order to employ the
physical scattering amplitude, we perform the on-shell reduction following
Starrett {\it et al.} \cite{Sta05} and assume that each amplitude is a function
of the effective velocity $v^{\pm}=\max(v_i^{\pm},v_f^{\pm})$ and momentum
transfer $s=|\Delta {\bf p}|$ linked to the scattering angle $\theta ^\pm $ by
$s=2v^\pm \sin (\theta ^\pm /2)$.

In view of what was said about the importance of the exchange interaction
between the electron and the target, let us first neglect the positron
contribution to the ampitude (\ref{eq:ampl}). The total differential
cross section for scattering from the state $a$ can then be written as
\[
\frac{d\sigma_a}{d\Omega}=4\sum_{b}\frac{v_{b}}{v_a}
\int g_{b}^*(\tilde {\bf q})g_{b}({\bf q})\bigl[f^-(\tilde v^{-},s)\bigr]^*
f^-(v^-,s)g_a^*(\tilde{\bf q}+\Delta{\bf p}/2)g_a({\bf q}+\Delta{\bf p}/2)
d^3{\bf q}d^3\tilde{\bf q},
\]
where $v_a$ and $v_{b}$ are the Ps velocities in the initial and final states.
We will assume now that the collision energy is well above a typical Ps
excitation threshold, so that we can neglect the dependence of $v_{b}$
and the momentum transfer $\Delta{\bf p}$ on $b$. Then the sum over $b$ 
yields $\delta({\bf q}-\tilde{\bf q})$ and we obtain
\begin{equation}
\frac{d\sigma_a}{d\Omega}=
4\int \big|f^-(v^-,s)\big|^2|g_{a}({\bf q}+\Delta{\bf p}/2)|^2 d^3{\bf q} .
\label{eq:dif_crs}
\end{equation}
If the state $a$ is the ground state of Ps, the function $|g_a|^2$ in
Eq.~(\ref{eq:dif_crs}) exhibits a sharp peak 
at ${\bf q}+\Delta{\bf p}/2\approx 0$, and can be replaced by the $\delta $
function in the ``peaking approximation'' \cite{Bri77}.
 Equation (\ref{eq:vv}) then gives
${\bf v}_i^-\approx {\bf p}_i/2={\bf v}_i$ and
${\bf v}_f^-\approx {\bf v}_i+\Delta {\bf p}$, where ${\bf v}_i$ is the
incident Ps velocity. Thus, we can neglect the variation of $f^-(v^-,s)$ when
integrating over ${\bf q}$ in Eq. (\ref{eq:dif_crs}), and obtain
\[
\frac{d\sigma_a}{d\Omega}=4|f^-(v^-,s)|^2 .
\]
For calculation of the integral cross section we note that the Ps and electron
solid angles are related by $d\Omega = d\Omega ^- /4$, which gives
\begin{equation}\label{eq:sigeq}
\sigma_a({\rm Ps})=\sigma_a(e^-).
\end{equation}
Hence the total integral cross sections for Ps-$A$ and $e^-$-$A$
collisions are equal for equal incident {\it velocities} $v^-\approx v_i$.

In deriving this result we made several approximations, assuming that the
collision energy is high compared to the typical Ps excitation energy and that
the $e^-$-$A$ interaction dominates Ps-$A$ collisions. The latter
assumption is supported by experiments and calculations showing that for
scattering from noble-gas atoms in the intermediate energy range
($v=0.5$--2 a.u.) the $e^+$-$A$
collision cross sections are much smaller than the $e^-$-$A$ cross sections
(see, e.g., \cite{Kauppila81,Dababneh82} and \cite{Kimura00} for molecules).
In contrast, at low energies ($<1$ eV) 
the $e^+$-$A$ cross sections are larger due to the effect of virtual Ps formation 
\cite{Dzu96,Sur05},
leading to larger absolute values of the scattering lengths.

We will present now the results for the partial and total cross sections of
Ps-Kr and Ps-Ar scattering obtained by full three-dimensional integration
in Eq. (\ref{eq:ampl}). The scattering phase shifts necessary for the
calculation of the electron and positron scattering amplitudes are taken from
polarized-orbital calculations of McEachran {\it et al.}
\cite{McE79,McE80,McE83,McE84}. Note that many calculations of $e^+$ collisions
with Ar and Kr have been published after the work of McEachran {\it et al.}
(see, e.g., \cite{Pov13} and references therein), some at a more advanced
level. However, our goal in the present work is to demonstrate the
correspondence between $e^-$-$A$ and Ps-$A$ scattering, and
Refs.~\cite{McE79,McE80,McE83,McE84} are most convenient for this purpose
since they present the scattering phase shifts calculated consistently for all
four cases ($e^-$-Ar, $e^+$-Ar, $e^-$-Kr, and $e^+$-Kr).

In the present work we have calculated the amplitudes and cross sections
for Ps elastic scattering and excitation of $n=2$ states of Ps. 
 To obtain the total cross section, the Ps
ionization (i.e., break-up) contribution should be added. This was taken from
the calculations of Starrett {\it et al.} \cite{Sta05} for the velocity range
between ionization threshold ($v=0.5$~a.u.) and $v=1.7$~a.u. For higher
velocities we use a smooth extrapolation.

\begin{figure}[ht]
\centering
\includegraphics[width=12cm]{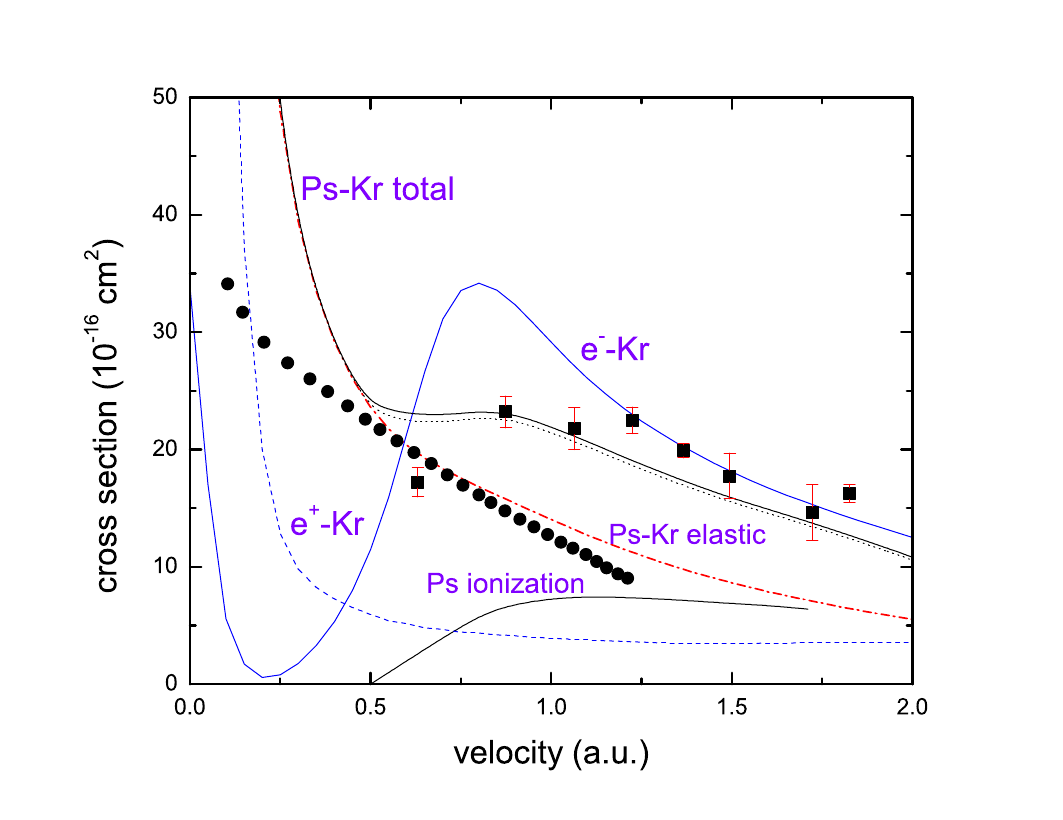}
\caption{(Color online) $e^-$-Kr, $e^+$-Kr and Ps-Kr scattering cross sections. Dotted
black line is the sum of elastic and ionization \cite{Sta05} cross section, 
and the line ``Ps-Kr total'' also contains contribution from excitation of the
$n=2$ levels of Ps. Solid red line ``SE" is
elastic cross section from static-exchange
calculations of Blackwood {\it et al} \cite{Bla02}. 
Experimental data with error bars are from
Ref.~\cite{Bra10}, and the data for $e^-$-Kr and $e^+$-Kr scattering are from
\cite{McE80,McE84}.} 
\label{fig:krypton}
\end{figure}

Fig. \ref{fig:krypton} shows the cross sections for Kr, plotted as functions
of the velocity of the projectile ($e^-$, $e^+$ and Ps). 
Cross sections  for $e^+$ at $v>1.3$ a.u. were obtained by extrapolation of the scattering phase shifts of McEachran {\it et al.} \cite{McE80}. 
By using different extrapolation 
schemes, we have estimated the uncertainty of the cross sections at 
$v>1.3$ a.u. to be less than 5\%.
The Ps-Kr elastic cross section dominates the total below the ionization
threshold at $v=0.5$ a.u., but above this velocity the ionization contribution
is substantial and becomes comparable to the elastic cross section
above $v=0.7$ a.u. As a result of the rise in the ionization contribution, the total cross section exhibits a weak maximum at $v=0.8$ a.u., 
much flatter than that in the $e^-$-Kr cross section, mainly because
the Ps elastic cross section grows rapidly towards lower energies. This growth
is mostly due to the $e^+$ contribution to the amplitude (\ref{eq:ampl}). In
contrast, for velocities above 0.5 a.u., $e^-$-Kr scattering dominates, and the
total Ps-Kr cross section approaches that of $e^-$-Kr scattering.

Comparison with the elastic cross section from the
static-exchange 
calculations of Blackwood {\it et al.}
\cite{Bla02} 
shows agreement with our Ps-Kr elastic cross section for $v=0.5$--1~a.u.
The impulse approximation is not expected to work at low collision energies 
below the Ps ionization threshold which suggests that the sharp upturn of 
the cross section below $v=0.5$~a.u. is an artifact. 
Instead the cross section should approach the zero-energy limit
of Mitroy and Bromley \cite{Mit03} $\sigma=(6$--$24)\times 10^{-16}$~cm$^2$
calculated by the stochastic variational method. (The numbers indicate the
bounds due to uncertainty in the input parameters of their model). It is
somewhat surprising that our results remain in good agreement with the
static-exchange calculations \cite{Bla02} down to the velocity 0.3 a.u. 
We should note, though, that because of the frozen-target approximation, the
van der Waals interaction is not effectively included in 
calculations~\cite{Bla02},
and in the case of Ps-H scattering it was shown by the same group 
\cite{Cam98,Bla02a} that inclusion of virtual excitations of Ps and 
the target leads
to much smaller cross sections in the low-energy region. Note also that the
Ps excitation cross section is very small compared to elastic and ionization,
in agreement with Blackwood {\it et al.} \cite{Bla02}.

Note that in order to obtain a peak in the total Ps cross section, adding the
inelastic contribution, mainly ionization cross section for Ps, is crucial. As a
result, our total Ps-Kr scattering cross sections agree well with the
measurements of Brawley {\it et al.} \cite{Bra10}, although, in contrast to
observations, the calculated peak is very weak, and looks more like an 
inflection point due to failure of the impulse approximation at low energies.

\begin{figure}[ht]
\centering
\includegraphics[width=12cm]{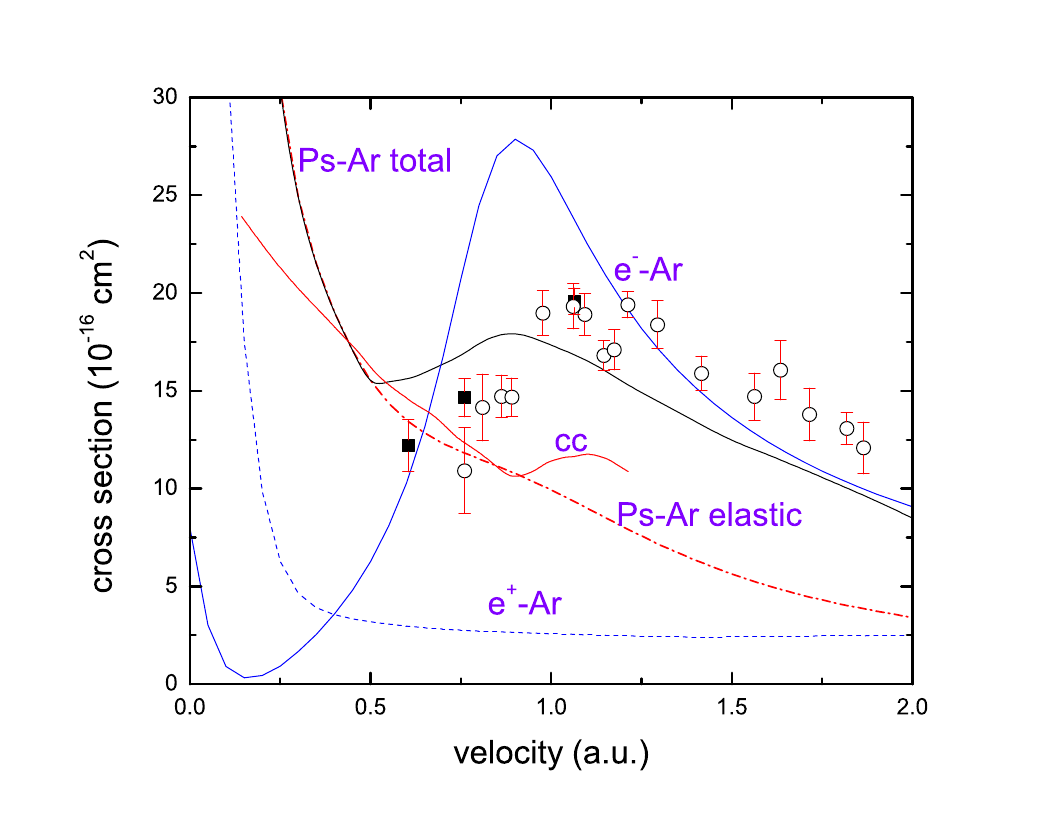}
\caption{(Color online) $e^-$-Ar, $e^+$-Ar and Ps-Ar scattering cross sections.
The line ``Ps-Ar total'' contains contribution of the elastic
scattering calculated from Eq.~(\ref{eq:ampl}) and ionization (from
Ref.~\cite{Sta05}). Solid red line ``cc": elastic cross sections 
from 
close-coupling calculations \cite{Bla02}. 
Circles with error bars: experiment \cite{Gar00}. Squares
with error bars: experiment \cite{Bra10}. The data for $e^+$-Ar and $e^-$-Ar
scattering are from \cite{McE79,McE83}. Cross sections for $e^+$ at
$v>1.3$ a.u. were obtained by extrapolation of the scattering phase shifts of
McEachran {\it et al.} \cite{McE79}.} 
\label{fig:argon}
\end{figure}

In Fig. \ref{fig:argon} we present the results for Ar. Since the
Ps excitation cross sections are very small compared to elastic scattering,
we include only elastic and ionization contributions in the total. The major
features in the elastic and total cross sections are the same as
for Kr, and the peak at $v=0.9$ a.u. in the total cross section is more
pronounced. The elastic cross section is
close to the results of the frozen-target close-coupling calculations \cite{Bla02} for $v=0.5$--0.8~a.u. (where the latter is dominated by elastic scattering), and the total agrees well with the experiment \cite{Gar00,Bra10}. However, as in the case of Kr, the upturn below
$v=0.5$~a.u. is an artifact of the impulse
approximation. Most likely, the
cross section below this velocity should approach the zero-energy limit of
Mitroy and Ivanov \cite{Mit02}, $\sigma=(7$--$16)\times 10^{-16}$ cm$^2$.
The sharp upturn in the Ps scattering cross section at low velocities is due to the contribution of the $e^+$-Ar scattering amplitude. When this contribution is neglected, the low-energy Ps scattering cross section becomes substantially lower and falls within the the Mitroy-Ivanov boundaries. The same is true for the Ps-Kr scattering. It is not clear whether this result is fortuitous or physically significant.

In conclusion
our work offers a clear physical and quantitative theoretical explanation
for the unexpected similarity between the Ps and electron scattering for 
equal projectile velocities uncovered recently by experiment 
\cite{Bra10,Bra10a}.
 Physically, this phenomenon occurs due to the relatively
weak binding and diffuse nature of Ps, and the fact that electrons 
scatter more strongly than positrons off atomic targets for incident velocities
$v\sim 1$ a.u. 
 Such similarity appears to be a generic phenomenon, 
and it is natural that an explanation is offered by using an approximation 
(in this case, impulse approximation) which emphasizes the physics of the 
problem. By contrast, large-scale numerical calculations for specific targets 
may be capable of reproducing experimental data but often lack the transparency
required for providing physical insight.
Note that the present impulse-approximation approach can also be improved
by extension of the $e^-$-$A$ and $e^+$-$A$ scattering amplitudes off the
energy shell and by considering higher-order approximations of the Faddeev
theory \cite{Fab92,Khr95}.

The authors are grateful to G. Laricchia for stimulating discussions and for
providing experimental data in numerical form.

\end{document}